# DYNAMIC AERODYNAMIC-STRUCTURAL COUPLING NUMERICAL SIMULATION ON THE FLEXIBLE WING OF A CICADA BASED ON ANSYS

DONG Qiang, ZHANG Xi-jin, ZHAO Ning

School of Mechatronics, Northwestern Polytechnical University, China
strongdq@aliyun.com

## ABSTRACT

*Most biological flyers undergo orderly deformation in flight, and the deformations of wings lead to complex fluid-structure interactions. In this paper, an aerodynamic-structural coupling method of flapping wing is developed based on ANSYS to simulate the flapping of flexible wing. Firstly, a three-dimensional model of the cicada's wing is established. Then, numerical simulation method of unsteady flow field and structure calculation method of nonlinear large deformation are studied basing on Fluent module and Transient Structural module in ANSYS, the examples are used to prove the validity of the method. Finally, Fluent module and Transient Structural module are connected through the System Coupling module to make a two-way fluid-structure Coupling computational framework. Comparing with the rigid wing of a cicada, the coupling results of the flexible wing shows that the flexible deformation can increase the aerodynamic performances of flapping flight.*

## 1. INTRODUCTION

Mimic-insect Flapping Wing Micro Air Vehicles (FWMAV) has broad application prospects [1], is currently a hot research field on MAV. Studies [2] have shown that most insect wings have significant flexible deformation during the flight, and some researches [3] have also found that the flexible deformation has a greater impact on the aerodynamics performance of flapping wings.

Heathcote et al [4] studied the effects of the spanwise flexibility on the wing thrust generated during the wing flight up and down by experiment, and the results shown that the flexible wings have higher thrust and propulsive efficiency than rigid wings. Combes and Daniel [5] examined the relative contributions of fluid-dynamic and inertial-elastic forces to passive wing bending, and found that the contribution of wing inertia to instantaneous wing shape is major compared to the contribution of aerodynamic loading.

Hamamoto et al [6] have performed fluid-structure interaction analysis based on the arbitrary Lagrangian-Eulerian method on a deformable dragonfly wing in hover, they found that the rigid wing need for greater peak torque and peak power than the flexible wing on generating the same lift. Tang et al [7] coupled linear finite element and aerodynamic Navier-Stokes solver based pressure to analyze the flexible wing in two-dimensional, and found that the deformation of the wing has very small impact on the thrust generating. Overall, the principle of the deformable flexible flapping wing not yet have a clear understanding, and there is not a three-dimensional nonlinear numerical CFD/CSD coupling method, so this area still needs further exploration.

In this paper, the three-dimensional model of cicada's wing is established, and ANSYS is used as a platform to explore the numerical simulation method of unsteady flow field based on Fluent and the nonlinear structure calculation method, and then the two are combined to simulate the dynamic aerodynamic-structural coupling of flapping wing, where the instantaneous aerodynamic forces are measured. The lift coefficient and trust coefficient are compared between the rigid wing and the flexible wing. The results provide some reference for the design and control of FWMAV.

## 2. MATHEMATICAL DESCRIPTION OF CICADA WING

### 2.1. Three-dimensional model of flexible Cicada wing

Cicada's wing is composed of veins and membranes (Figure 1), and the membranes are divided into many tiny pieces by the veins. On establishing the three-dimensional model of the wing, dues to the actual production and simplify the calculation, the structure of the cicada's wing has been simplified, where the 3-D model of the veins and membranes are established separately, then the two models are bonded to form the whole model of cicada's wing.

For the model of the wing membranes, the high-definition cameras are used to capture the true picture of cicada's wings, then the vector image is imported to Pro/E, the edge contour of cicada's wings is plotted (here the relative movement between the forewing and underwing of cicada is ignored), and the contour line is drew only little thickness to obtain the three-dimensional simplified model of the membranes, shows in Figure 1, wing length = 52.5mm, wing wide = 27.0mm, membranes thickness = 100um.

For the model of the vein structure, the main veins which play the major role in supporting the wings are selected (Figure 1). The cross-section of the veins is rectangular, which could be conjoint well with the membranes, and increases gradually in width and thickness from the wing tip to wing root ,these veins are fixed in the wing root to form a whole vein model.

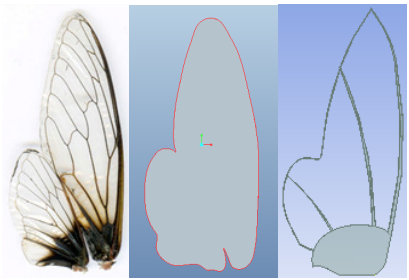

Figure 1. Cicada wing and its 3D model

According to most productions of flapping-wing aircraft, the carbon fiber composite and polyimide film are selected as the materials of veins and membranes, respectively. So that the 3-D model of the established wing will have a certain elevated flexibility.

## 2.2. Motion equations of cicada's wing

Inertial reference system OXYZ is established, which is also known as fixed coordinate system. The body-fixed coordinate system *oxyz* is established on the wing, where the *x*-axis is along the wing chord, the *y*-axis is along the wing span. Inertial coordinate system remains stationary, while the body-fixed coordinates is fixed on the wing, and moves with the wing. In the initial (*t*=0), the body-fixed coordinates system is coincided with the inertial coordinate system. Figure 2 shows the movement of Cicada's wing. The simplified motion of cicada's wings is obtained according to the in-house experiments, namely, the wing is flapping up and down around the Y-axis in the fixed coordinate system, while rotating around the x-axis at the upper and the lower vertex of flapping, so that the wing in down-stroke remains 0 degrees of attack, and in up-stroke remains 15 degrees of attack. In Figure 4, $\theta$ is the angle between the plane of the flapping and the horizontal surface, $\beta$ is the twist angle.

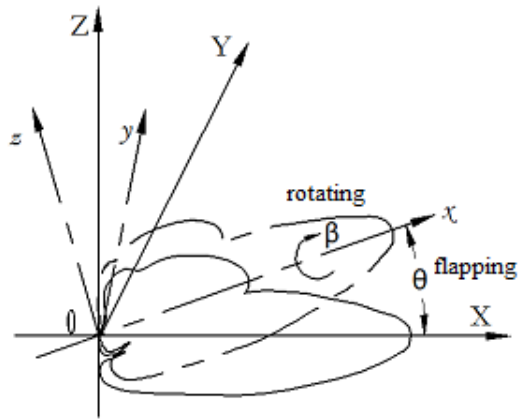

Figure 2. Movement of the cicada wing

The movement of cicada's wings can be expressed by the following two equations:

1) Flapping around Y-axis in fixed coordinate system;

$$\theta(t) = \theta_0 \sin(2\pi f t) \qquad (1)$$

2) Rotating around *x*-axis in body-fixed coordinates system;

$$\beta(t) = \begin{cases} \beta_0 & (0 \leq t < 19T/80) \\ 0.5\beta_0 \sin[40\pi f(t-18T/80)] + 0.5\beta_0 & \\ & (19T/80 + nT \leq t < 21T/80 + nT) \\ 0 & (21T/80 + nT \leq t < 59T/80 + nT) \\ 0.5\beta_0 \sin[40\pi f(t-56T/80)] + 0.5\beta_0 & \\ & (59T/80 + nT \leq t < 61T/80 + nT) \\ \beta_0 & (61T/80 + nT \leq t < 99T/80 + nT) \end{cases} \qquad (2)$$

Where *f* is the flapping frequency (Hz), $\theta_0$ is flapping amplitude (rad), $\beta_0$ is the maximum torsion angle (rad), *T* = 1/*f* is flapping cycle.

## 3. NUMERICAL ANALYSIS

### 3.1. Computation of unsteady aerodynamics

The Reynolds number of flapping wing flight is about $10^2 \sim 10^4$, which belongs to the low Reynolds number unsteady air flow. Currently, although the calculation method for this problem has well developed, but more of them are complex, and with the development of commercial CFD software, it could be calculated simply, There ANSYS Fluent is used to explore the simulation about the unsteady flow of flapping wing on three-dimensional.

The control equations for calculating is the three-dimensional incompressible N-S equations. In inertial coordinate system OXYZ, the three-dimensional incompressible N-S equations [8] as follows:

$$\frac{\partial u}{\partial x}+\frac{\partial v}{\partial y}+\frac{\partial w}{\partial z}=0 \qquad (3)$$

$$\frac{\partial u}{\partial t}+u\frac{\partial u}{\partial x}+v\frac{\partial u}{\partial y}+w\frac{\partial u}{\partial z}=-\frac{\partial p}{\partial x}+\frac{1}{Re}\left(\frac{\partial^2 u}{\partial x^2}+\frac{\partial^2 u}{\partial y^2}+\frac{\partial^2 u}{\partial z^2}\right) \qquad (4)$$

$$\frac{\partial v}{\partial t}+u\frac{\partial v}{\partial x}+v\frac{\partial v}{\partial y}+w\frac{\partial v}{\partial z}=-\frac{\partial p}{\partial y}+\frac{1}{Re}\left(\frac{\partial^2 v}{\partial x^2}+\frac{\partial^2 v}{\partial y^2}+\frac{\partial^2 v}{\partial z^2}\right) \qquad (5)$$

$$\frac{\partial w}{\partial t}+u\frac{\partial w}{\partial x}+v\frac{\partial w}{\partial y}+w\frac{\partial w}{\partial z}=-\frac{\partial p}{\partial z}+\frac{1}{Re}\left(\frac{\partial^2 w}{\partial x^2}+\frac{\partial^2 w}{\partial y^2}+\frac{\partial^2 w}{\partial z^2}\right) \qquad (6)$$

Where u, v and w are fluid velocity along X-axis, Y-axis and Z-axis, respectively. p is pressure, t is time. In the above equations, Re=$cU/v$, is the Reynolds number, where $v$ is the kinematic viscosity coefficient of air, $U$ is the average linear velocity of flapping wings, $c$ is the mean chord of the wings.

The ICEM CFD software is used for meshing the fluid grid, and the grid type is unstructured tetrahedral. On account of the large deformation area, the dynamic moving grid skills which combine the spring-based smoothing and the local remeshing method are taken to update grid.

For unsteady flow, the renormalization group (RNG) k-ε turbulence model is chose, the finite volume method is used to separate the computing areas, and the discrete equations are solved by the pressure-velocity coupling method –SIMPLE.

The lift and thrust coefficients can be expressed as follows:

$$C_L = \frac{F_L}{0.5\rho U^2 S} \qquad (7)$$

$$C_T = \frac{F_T}{0.5\rho U^2 S} \qquad (8)$$

Where $F_L$, $F_T$, $\rho$, $U$, $S$ are lift, thrust, air density, average line speed of flapping wings, wing area, respectively.

In order to compare and analysis expediently, the mean of the lift coefficient and thrust coefficient of a cycle are used in this paper.

### 3.2. Analysis of structure dynamics

The flapping of a wing is a transient motion which varying with time, structural calculations use the transient dynamics module (Transient Structural) in ANSYS. from classical mechanics, the dynamics universal equation of an object as follows:

$$[M]\{x''\} + [C]\{x'\} + [K]\{x\} = \{F(t)\} \tag{9}$$

where $[M]$ is the mass matrix; $[C]$ is the damping matrix; $[K]$ is the stiffness matrix; $\{F(t)\}$ is all the external force vector which putting on the object; $\{x''\},\{x'\},\{x\}$ are the acceleration vector, velocity vector and displacement vector of the moving object, respectively.

Due to the large deformation and displacement of the wings in flapping, the nonlinear structure calculation is taken.

### 3.3. Coupling method [9], [10]

The calculation of flow-solid coupling follows basic conservation principles at the interface, it satisfies the equal and conservation of the force (F) and displacement (d) between fluid and solid.

$$\begin{cases} F_f = F_s \\ d_f = d_s \end{cases} \tag{10}$$

Where the subscript of $f$ represents fluid and $s$ represents solid.
For data transfer at the coupling interface, multi-field solver-MFX uses conservative interpolation method, which combines pixel concept, unit split, build new control surfaces, bucket algorithm and other ways to complete the data transfer, making the parameter data transfer accurately from global to local at the interface.

Convergence conditions: in each time step, the value of the transferred data in continuous iterative calculation is φ (force or displacement), in all the transfer position $l$, the value of the transferred data in current iterative calculation $\varphi_l^{curr}$, in previous iteration calculation $\varphi_l^{pre}$, the difference between $\varphi_l^{curr}$ and $\varphi_l^{curr}$, relatives to the average value of transferred data through all position on the current iteration, is defined as $\overset{\wr}{\Delta}_l$, the root mean square value of $\overset{\wr}{\Delta}_l$ is RSM, namely:

$$\begin{cases} \overset{\wr}{\Delta}_l = \dfrac{\left(\varphi_l^{curr} - \varphi_l^{pre}\right)/\omega}{0.5 \times (\max|\varphi| - \min|\varphi| + \overline{|\varphi|})} \\ \text{RSM} = \sqrt{\overline{\left(\overset{\wr}{\Delta}_l\right)^2}} \end{cases} \tag{11}$$

There takes $RSM_{max} \leq 10^{-14}$ as the convergence condition, where ω is the under relaxation factor.

### 3.4. Coupling process

ANSYS provides the two-way fluid-structure interaction module-System Coupling, which uses the orderly coupling method. Figure 3 gives the calculation process.

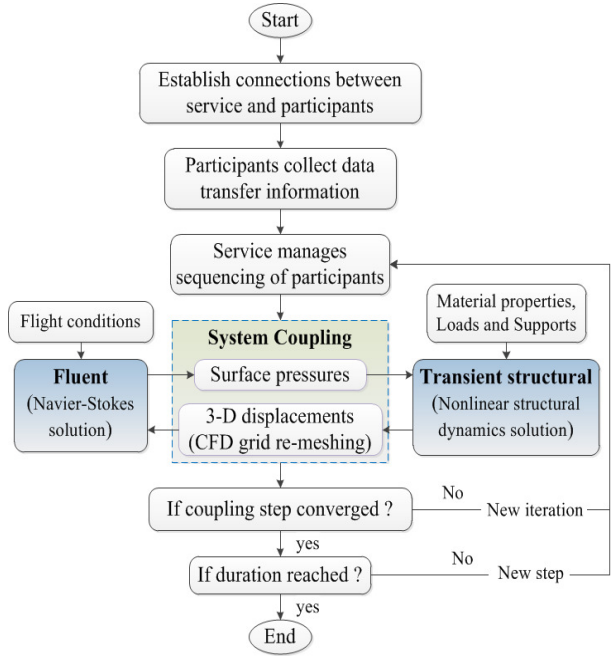

Figure 3. Process of two-ways fluid-structure interaction

The specific process as follows:

(1) The connection among Fluent module, Transient structural module and System Coupling module is established, and the coupling calculation parameters are set;
(2) System Coupling module controls the calculated progress, Transient structural module receives the aerodynamic force of previous calculated step on the flapping wing surface, then solves the structural dynamics equations, exports the displacement of each grid point on flapping wings at this time;
(3) Fluent module receives each grid point displacement, uses the dynamic mesh method to update the deformation of flexible flapping wing, then solves the unsteady Navier-Stokes equations, and exports the unsteady aerodynamic force of the flapping wing surface;
(4) Checking the convergence condition, if RSM is converged, calculating the next time step. Otherwise, repeating step (2) to step (3).

## 4. COMPUTATION VALIDATION

### 4.1. CFD validation by a case of rigid wing

In order to verify the validity of the CFD results, the flapping motion of Drosophila model which was used in the experiment by Dickinson et al [11] is simulated. In their experiment, the wing length is 0.25 m and the wing area is 0.0167m², the fluid density is $0.88 \times 10^3$ kg m$^{-3}$, the flapping

International Journal of Recent advances in Mechanical Engineering (IJMECH) Vol.3, No.4, November 2014

frequency is 0.145Hz, the flapping amplitude is 160°, the angle of attack in flapping up and down are 40°, and the rotation phase advances 8% than the flapping phase. For comparison, the UDF function is used to calculate the lift of Drosophila model in flapping directly (Fluent have mutations in the beginning of the calculation, so the stable calculated data is taken). The comparison between the simulation results and experimental results of Dickinson et al. is shown in Figure 4, and the trend of the two are corresponded well, indicating that the use of Fluent to simulate the unsteady flow field of flapping wing on three-dimensional is feasible.

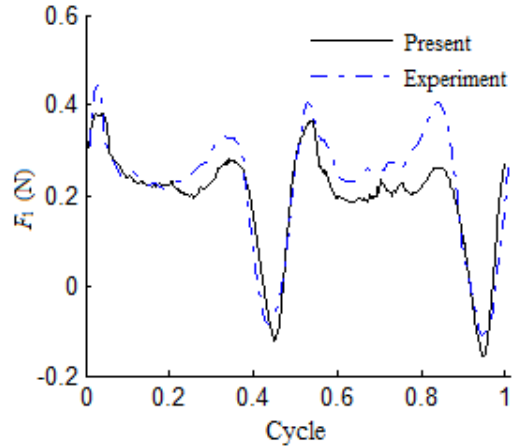

Figure 4. Comparison of computed results and experiment data from Dickinson.

### 4.2. Validation of structure computation

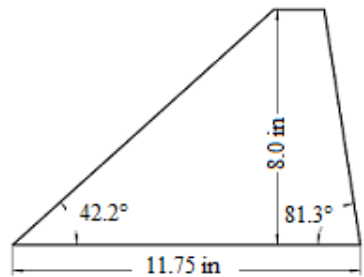

Figure 5.The delta wing model of cut tip

According to the literature [12] the delta wing model (Figure 5) of cut tip is used to conduct a nonlinear static analysis, the material parameters: $E = 2.07 \times 10^{11}$Pa, $u = 0.25$, thickness is 0.888mm, fixing the roots, and putting a concentrated load $P_Z = 10$N on the endpoint of the trailing edge. The calculation and the literature results of the load point displacement are listed in Table 1, both are very approximate, and the displacement of the tip approximately 16 times the thickness, reflecting the structural large deformation obviously.

Table 1. Displacement of the load point

| $P_Z$=10N | Displacement |
|---|---|
| MSC. NASTRAN computation | 13.9mm |
| Literature result | 13.8mm |
| Present computation | 13.86mm |

31

## 5. ANALYSIS AND DISCUSSION OF THE NUMERICAL RESULTS

The actual shape of an insect wing is thin, and the wing surface is uneven. Rees [13] thought that, from the aerodynamic point of view, the function of the wing structure is precisely similar with a airfoil of slight thickness. Luo Guo-yu [14] used a numerical simulation to validate that this uneven structure did not affect the overall aerodynamic performance of flapping motion. So as to simplify the calculation, the wing surface structure's effect on aerodynamic performance is ignored here, and for coupling calculation, the wing membrane is acted as the interface in fluid-structure interaction. Namely, in structural calculation, both the veins and membranes are taken into account to compute the deformation and displacement of wing membrane, then the date is transferred to the flow field (only considering the wing membrane) to update the membrane's shape, Fluent module calculated the aerodynamic force of the membrane surface, and transferred it to the structure module. Then the System Coupling module looped the iteration.

According to the characteristics of cicada wings in flight, flapping frequency $f$=40Hz, torsion angle $\beta_0$=15°, flapping amplitude $\theta_0$=90°, the time step is 0.0001s, flow velocity is $U_0$ = 5m/s (the flight velocity of cicada is 2~8m/s, we used the average velocity). After calculating 250 time steps, the instantaneous lift coefficient and thrust coefficient of the flexible wing and rigid wing are measured and compared in figure 5, where the non-dimensional time from 0.13 to 0.63 is downward stroke, and others are upward stroke, the wing rotating around the upper vertex (t=0.13) and the lower vertex (t=0.63) of flapping. Table 1 shows the mean lift coefficient and mean thrust coefficient of flexible wing and rigid wing in a period.

In Figure 6(a), the flexible wing generates higher positive lift peak than rigid wing in the rotation process (t=0.13 and t=0.63) and the downward stroke (t=0.13~0.63) during the flight. In the upward stroke, the flexible wing generates a small up peak. Combining with Table 2, where the mean lift coefficient of flexible wing almost doubles that of rigid wing, we draw the conclusions that the flexible deformation of the flapping wings can increase the lift generated on the wings during the flight.

Figure 6(b) shows that, in the downward stroke (t=0.13~0.63), the thrust coefficient of flexible wing fluctuates around the thrust coefficient of rigid wing, and the mean thrust coefficient of flexible wing within downstroke a little higher than that of rigid wing. In the upward stroke, obviously, the flexible wing shows better thrust peak compared to rigid wing. The mean thrust coefficient of flexible wing almost doubles that of rigid wing (Table 2). That is to say, the flexible deformation of the flapping wings can increase the thrust on the upward stroke.

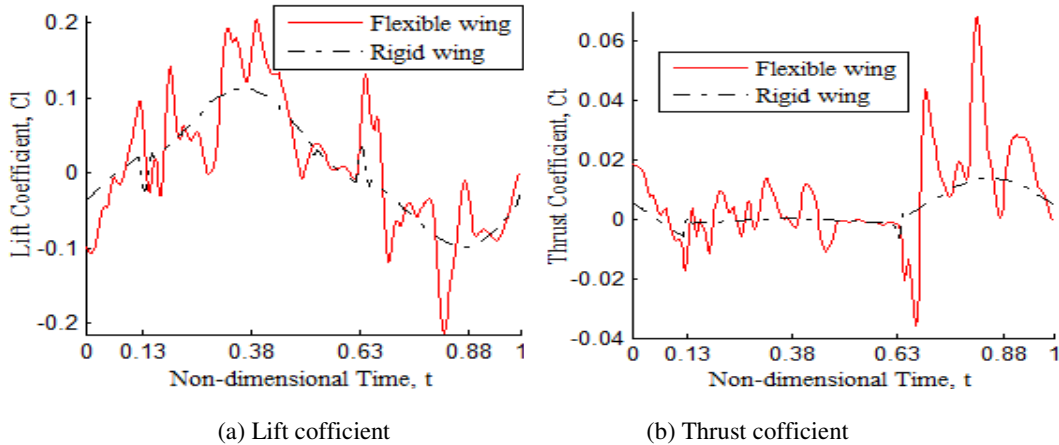

(a) Lift cofficient          (b) Thrust cofficient

Figure 6. Comparison of lift and thrust coefficients between rigid wing and flexible wing

Table 2. Mean lift and thrust coefficient the of rigid wing and flexible wing in a period

|  | Mean lift coefficient | Mean thrust coefficient |
|---|---|---|
| Rigid wing | 0.0064 | 0.0030 |
| flexible wing | 0.0121 | 0.0067 |

## 6. CONCLUSIONS

In this paper, a dynamic aerodynamic-structural coupling method of flapping wing was developed based on ANSYS, and the numerical simulations on the flexible wing and rigid wing of a cicada is conducted.

The results show that, the flexible deformation of the wing (established in this paper) can increase the lift coefficient peak in the rotation process and the downward stroke, and it also increases the thrust coefficient peak in the upward stroke. Finally, the mean lift coefficient and mean thrust coefficient of flexible wings both almost double that of rigid wings.

## ACKNOWLEDGEMENTS

This work is supported under National Natural Science Foundation of China (51175426) and Natural Science Basic Research Plan in Shaanxi Province of China (2012JM7010). The authors would like to thank Prof. Zhao Ning and A.P. Zhang Xi-jin for his work in help and instruction.

## AUTHORS

**Dong Qiang**

Graduate Student, School of Mechatronics, Northwestern Polytechnical University, China.
Field of research: Design and Analysis of Flapping Wing Micro Air Vehicles (FWMAV)

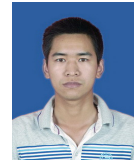